\documentclass[aps,prb,twocolumn,groupedaddress]{revtex4}
\usepackage{graphics}
\usepackage{bm}
\usepackage{epsfig}

\newcommand{\be}{\begin{equation}}
\newcommand{\ee}{\end{equation}}
\newcommand{\bea}{\begin{eqnarray}}
\newcommand{\eea}{\end{eqnarray}}

\newcommand{\ua}{\uparrow}
\newcommand{\da}{\downarrow}

\begin{document}

\title{Supercurrent induced domain wall motion}

\author{P. D. Sacramento$^1$, L. C. Fernandes Silva$^1$, G. S. Nunes$^{1,2}$, M. A. N. Ara\'ujo$^{1,3}$ and V. R. Vieira$^1$}
\affiliation{$^1$ CFIF, Instituto Superior
T\'ecnico, TU Lisbon, Av. Rovisco Pais, 1049-001 Lisboa, Portugal}

\affiliation{$^2$ Department of Engineering and Management, ISCTE, Avenida das For\c{c}as Armadas, 1600 Lisboa, Portugal
}

\affiliation{$^3$ Departamento de F\'{\i}sica, Universidade de \'Evora, P-7000-671, \'Evora, Portugal}

\date{\today}

\global\long\def\ket#1{\left| #1\right\rangle }

\global\long\def\bra#1{\left\langle #1 \right|}

\global\long\def\kket#1{\left\Vert #1\right\rangle }

\global\long\def\bbra#1{\left\langle #1\right\Vert }

\global\long\def\braket#1#2{\left\langle #1\right. \left| #2 \right\rangle }

\global\long\def\bbrakket#1#2{\left\langle #1\right. \left\Vert #2\right\rangle }

\global\long\def\av#1{\left\langle #1 \right\rangle }

\global\long\def\tr{\text{Tr}}

\global\long\def\im{\text{Im}}

\global\long\def\re{\text{Re}}

\global\long\def\sign{\text{sign}}

\begin{abstract}
We study the dynamics of a magnetic domain wall, inserted in, or juxtaposed to, a conventional
superconductor, via the passage of a spin polarized current through a FSF junction. 
Solving the Landau-Lifshitz-Gilbert
equation of motion for the magnetic moments we calculate the velocity of the domain wall
and compare it with the case of a FNF junction. We find that in several
regimes the domain wall velocity is larger when it is driven by a supercurrent.
\end{abstract}


\maketitle

\section{Introduction}

Spintronics has attracted considerable attention both from the fundamental point of view and from
the point of view of applications \cite{rmp1,fert}. 
Related to applications in magnetic registers, the manipulation of magnetic domain
walls by spin polarized currents has been a topic of interest, as opposed to direct
manipulation by application of local magnetic fields. 
Domain walls are present in magnetic materials and thin magnetic films. Their function is
to lower the magnetostatic stray field energy \cite{zabel}.  The basic phenomena of the DW motion
occur in a submicron-size ferromagnetic stripe \cite{nam}.
The manipulation of the location of
the domain walls results from the action of spin torques on the magnetic
moments \cite{ralph}, due to a polarized current \cite{slonczewski,berger,wang}.
This has been analyzed theoretically \cite{tatara,dugaev1,dugaev2} and experimentally
\cite{berger1,dws}. 
Typically the magnetic domain walls are inserted in magnetic semiconductors but interest
in other types of structures has arisen lately involving superconductors \cite{linders}.

We have been interested in the possible interplay between magnetic moments and superconductors \cite{satori,sakurai},
specifically the possibility of ordered magnetic moments in the superconductor or in its
vicinity \cite{us}. Magnetism and superconductivity typically compete and, therefore, most heterostructures
considered tend to separate the ferromagnetic and superconducting regions (often separated by
an insulator to prevent proximity effects). The presence of randomly located magnetic impurities
in a conventional superconductor destroys the superconducting order for 
small impurity concentrations \cite{abrikosov}
but we have shown \cite{rapid,jpcm2010} that, if the magnetic moments are correlated, 
superconductivity is much more
robust and prevails for much higher concentrations. Therefore, if the magnetic
moments are somewhat diluted we expect that superconducting order should remain. 
We will consider here diluted magnetic moments.

The effects of the polarization of the electrons due to the magnetic moments is well known to
be very local. Therefore, one may expect that passing a spin polarized current
through the superconductor (as in a normal metal) will induce a spin interaction between
the magnetic moments and the spins of the conduction electrons. In a conventional superconductor,
with singlet pairing, the spin density dies out inside a clean superconductor at a distance
of the order of the coherence length. Therefore,
in the case of an infinite system one does not expect a spin torque on the magnetic moments 
due to the incident spin polarized current.
However, the size of the nanostructure may be comparable to
the coherence length and, therefore, the spin density may not die out inside the superconductor.
Also, due to the polarization effect of the local magnetic moments on the spins of the conduction
electrons, a smaller decay of the spin density occurs. We have shown recently \cite{torquesc} that there is
an appreciable spin torque induced on the magnetic moments that is comparable to the one observed
in a normal metal. Clearly, one expects that considering system sizes of the order of the coherence
length will enhance the effect. Also, spin triplet superconductors will circumvent this issue altogether.

It is therefore interesting to further explore a junction of the type ferromagnet-superconductor-ferromagnet
(FSF) 
to find regimes where the domain
wall motion induced by the passage of the supercurrent may be more effective in the FSF junction
with respect to the ferromagnet-normal-ferromagnet (FNF) junction. Here we study the motion of the domain wall
solving the Landau-Lifshitz-Gilbert (LLG) equations for the magnetic moments. We find that
often the velocity of motion of the domain wall is actually larger for the supercurrent, as compared
to a normal metal.

A spin polarized current leads to an accumulation of spin density which interacts with the local
moments inducing a torque and consequent rotation. 
This leads ultimately to the intended shift of the
domain wall position by methods involving currents. 
These are more efficient and fast as compared to the
application of local magnetic fields to flip the magnetic moments.
For recent reviews see Refs. \cite{beach,yaroslav} and references therein.
Due to the finite size of the systems considered the motion of the domain
wall is limited by the system size. When the domain wall approaches the
boundary it gets distorted and stops its motion. We are therefore interested
here in studying the early time regimes during which the domain wall
is set into motion and has not yet distorted appreciably.

\section{Model}

In this work we consider a junction of either FNF or FSF types. 
For simplicity we consider a one-dimensional model system which is
a good description of a narrow superconducting wire.
Inside the superconducting region we place local magnetic
fields. On the left-hand side the ferromagnet exchange field, $h$, points in the $z$-direction and
on the right hand side the exchange field points in the $-z$ direction.  We take the domain wall
resulting from the diluted magnetic moments inserted in the superconductor 
centered at the midpoint of the
SC (see Figs. 3,4 below for $t=0$).
We use a lattice formulation for a 1D model system oriented along the $x$ axis, with Hamiltonian:
\be
\hat{H} = \hat{H}_c + \hat{H}_{c-S} + \hat{H}_S
\ee
where
\begin{eqnarray}
\hat{H}_c &=& - \sum_{n,\sigma} \left( \hat c_{n\sigma}^\dagger\hat c_{n+1\sigma} 
+ {\rm H.c.} \right) \nonumber \\ 
&+& U\sum_{n,\sigma} \left(\delta_{n,N_{SL}}+\delta_{n,N_{SR}} \right) 
\hat c_{n\sigma}^\dagger\hat c_{n\sigma} \nonumber\\ 
&+&
\sum_n \left( \Delta_n  \hat c_{n\ua}^{\dagger}\hat c_{n\da}^{\dagger} + {\rm H.c.}\right) 
\end{eqnarray}
is the electronic part of the Hamiltonian,
$\sigma=\ua,\da$ denotes the spin projections along the $z$ axis, 
we set the hopping to unity and, thereby set the energy scale, and we choose
the chemical potential to be zero. 
\be
\hat{H}_{c-S} = - \sum_{n,\sigma,\sigma'} J\bm{S}_n\cdot \bm{s}_n
\ee
is the interaction between the spin density of the conduction electrons and the impurity spins, with
$\bm{s}_n=\bm{\sigma}_{\sigma\sigma'} 
\hat c_{n\sigma}^\dagger \hat c_{n\sigma'}$
where $\bm{\sigma}=(\sigma^x,\sigma^y,\sigma^z)$ are the Pauli matrices, 
we assume the local magnetic moments $\bm{S}_n$ to be of unit length and 
to be Zeeman coupled to the electrons in the superconductor, behaving  as 
local magnetic fields $J\bm{S}_n\equiv \bm{\tilde J}_n$ in the superconductor 
and
\be
\hat{H}_S=-\sum_n J_{ex} \bm{S}_n\cdot \bm{S}_{n+1} + \frac{k_y}{2} \sum_n \left( S_n^y \right)^2
\ee
It is convenient to introduce a planar
anisotropy $k_y$, which for positive $k_y$ favors a state where the spins
are in the $x-z$ plane (the plane of the initial domain wall). 
At the superconductor interfaces with the
F or N systems ($N_{SL},N_{SR}$) we introduce a potential term, $U$, that
simulates the interface disorder \cite{ting}.
The local moments are distributed evenly inside the superconducting region and
interact with each other via a nearest-neighbor ferromagnetic
interaction $J_{ex}$. 
The superconducting region 
includes the magnetic moments that constitute the domain wall, as shown in Fig. 1 of 
Ref. \cite{torquesc}. 
Often we take the number of sites $N=160$,
and $N_{SR}-N_{SL}=100$.
We are considering for simplicity a singlet s-wave superconductor.

\begin{figure}
\includegraphics[width=0.45\textwidth]{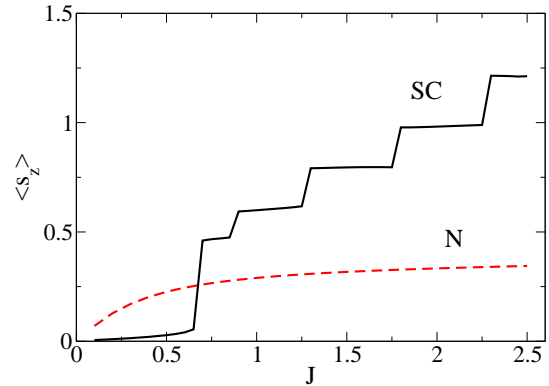}
\caption{\label{fig1} Evolution of $< s_z^l >$ as a function of $J$.
The various plateaus are obtained through quantum phase transitions. 
}
\end{figure}

\begin{figure*}
\includegraphics[width=0.99\textwidth]{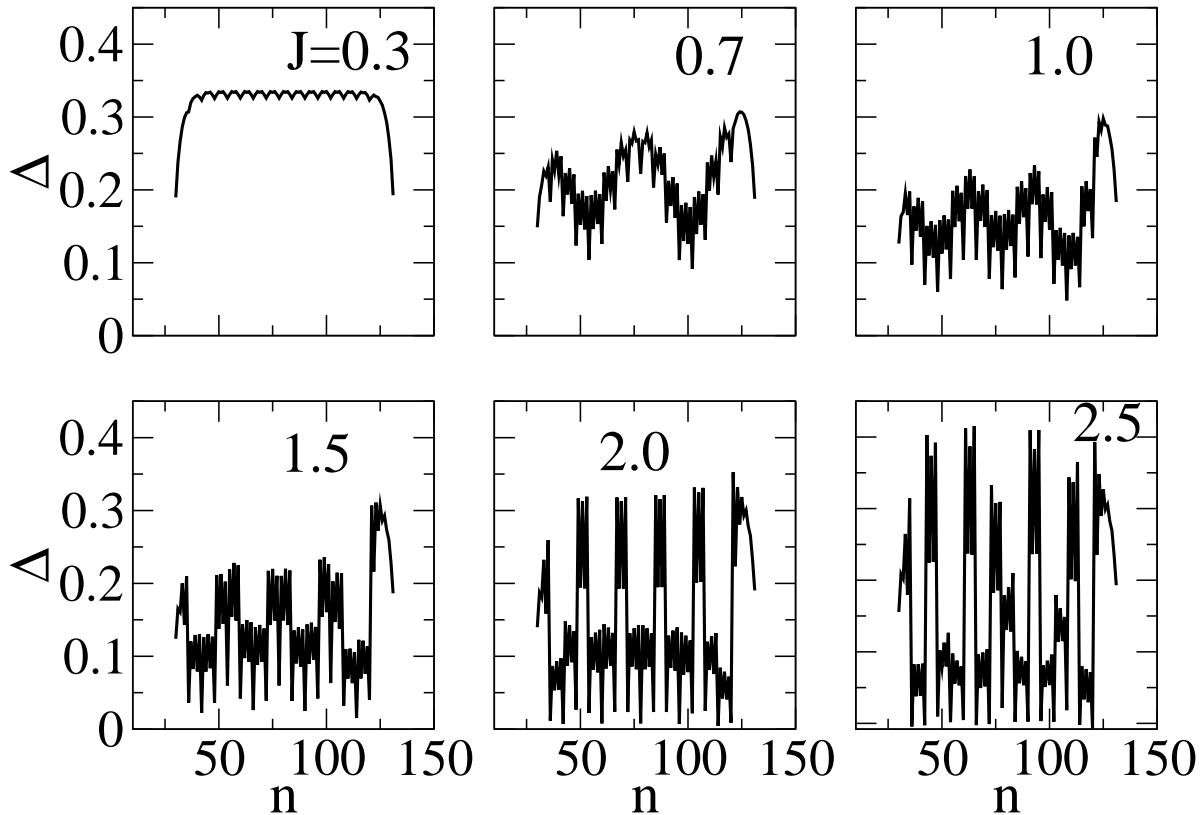}
\caption{\label{fig2} Spatial variation of the gap function for different values of $J$
from $J=0.3, \cdots, 2.5$.
Note that the number of peaks increases by one as the average magnetization jumps from
one plateau to the next.
}
\end{figure*}

The Bogoliubov-de Gennes (BdG) equations \cite{degennes} for $\hat{H}_c+\hat{H}_{c-S}$ determine
both the equilibrium electronic states and the scattering states when a current is passing through.
Their solution provide the wave functions in the usual way (see for instance \cite{us}).
The BdG equations
are solved self-consistently to find the profile of the gap function inside 
the superconductor,  $\Delta_n$.
The domain wall shape is calculated self-consistently in mean-field \cite{us}
ensuring that the energy is minimized and a vanishingly small equilibrium torque.
The procedure leads to a domain wall where the magnetization direction
interpolates between the $z$-direction on
the left hand side and the $-z$ direction in the right hand side, according to the 
orientations of the exchange fields in the two ferromagnets.
The procedure is described in detail in Ref. \cite{ting} and in Ref. \cite{torquesc}.

Using the transfer matrix method we determine the wave functions
at every site in the heterostructure and calculate the various physical quantities such
as the local spin densities, the spin and charge currents and the spin torque at
each site.
The torque at point $n$ is defined by the cross product between the magnetic moments
and the local spin density, $\bm{s}_n$,
\be
\label{tdef}
\bm{\tau}(n) = 2 J \bm{S}_n \times \bm{s}_n .
\ee
Assuming spin conservation the spin torque may also be calculated by the
difference in spin currents as
\be
\label{tdef2}
\tau_{\beta}(n)=j_{\beta}^s(n-1)-j_{\beta}^s(n) \,,
\ee
where $\beta=x,y,z$. 
The expressions for the spin currents are given 
in Ref. \cite{torquesc}.


In experiments, a potential difference, $V$, is imposed between the two sides 
of the heterostructure in a standard
way \cite{blonder,maekawa,lambert} leading to a current that moves from left to right. 
Imposing a potential difference at the ends of the heterostructure leads to an overall
torque on a given magnetic moment, that is the result of an integral over the incident energies,
up to the applied potential \cite{torquesc}. This torque leads to the motion of the magnetic moments, governed
by the Landau-Lifshitz-Gilbert equation \cite{LLG}.

\section{Equilibrium properties of the heterostructure}

We begin by considering a situation with no current flowing through the
system. In this case the BdG equations yield the equilibrium solution.
In Fig. \ref{fig1} we show the average magnetization of the electrons 
over the left half of the superconductor, $<s_z^l>$, 
along the $z$-direction, induced by the coupling to the local spins,
as a function of $J$. 
We consider $15$ spins distributed evenly in the superconductor (density $0.15$).
We see in the superconducting case that, as $J$ increases, the magnetization
increases, but at some points it changes discontinuously between various plateaus.
These discontinuities are due to quantum phase transitions in the system \cite{sakurai,us}.
In the case of one impurity and for small $J$ the average magnetization vanishes. 
Even though the local magnetic moment
polarizes the spin density of the electrons at the impurity location, 
the spin density of the electrons has fluctuations
(like Friedel oscillations) that compensate the perturbation. 
Above a certain critical $J_c$ the system is no longer
able to shield the perturbation and the overall magnetization jumps discontinuously to $1/2$. 
Various other discontinuities
occur at this first order quantum phase transition due to a level crossing, 
such as the gap function changing sign
at the impurity location, various entanglement \cite{entanglement} 
measures and the partial state fidelity \cite{fidelity}.
Increasing the number of magnetic impurities the discontinuities are still present, 
but at very small $J$ the magnetization
is not strictly zero \cite{us}, as shown in Fig. \ref{fig1}. 
We also show in Fig. \ref{fig1} the magnetization for
the normal case. We see that there are no discontinuities and the magnetization changes smoothly.
At small $J$ the magnetization is smaller in the superconducting case. As the first transition to a plateau occurs,
in the superconducting case, the magnetization becomes larger in the superconducting case.

The profile of the gap function is shown in Fig. \ref{fig2} for various spin couplings,
$J$. At small $J$ the gap function is approximately constant over the superconducting
region. It vanishes outside the superconductor and it has very small fluctuations
due to the finite size of the SC region and the finite number of impurity spins.
As the coupling grows the fluctuations in the gap function increase considerably
and are modulated by oscillations that increase in number as $J$ grows.
Note that, interestingly,
the number of peaks in the oscillations of the gap function as a function 
of space along the chain is related to the
various plateaus of the total magnetization. As one goes from one plateau to 
the next the number of peaks increases by one.
This is reminiscent of a LOFF \cite{loff} state where the gap function
has real space oscillations due to the finite momentum of the Cooper
pairs resulting from the splitting of the up and down spin bands
due to a magnetic field.
We recall that in a superconductor with 
magnetic impurities, if $J$ is large
enough, the gap function changes sign at the impurity locations. 
Here, however, we have an heterostructure. Interestingly,
we find that even though the gap function decreases significantly at the 
impurity locations as $J$ grows, it only changes
sign at very large $J\sim 2.5$. This is in contrast with an infinite 
superconductor where the gap function changes sign
at the location of the first quantum phase transition. 
This is shown in Fig. \ref{fig2}.

\section{Landau-Lifshitz-Gilbert equation and domain wall motion}

\begin{figure*}
\includegraphics[width=0.45\textwidth]{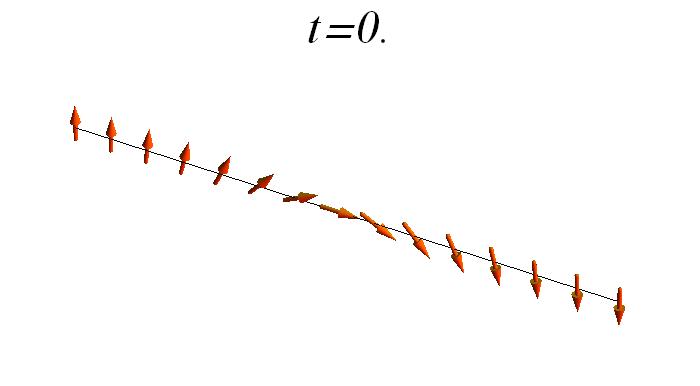}
\includegraphics[width=0.45\textwidth]{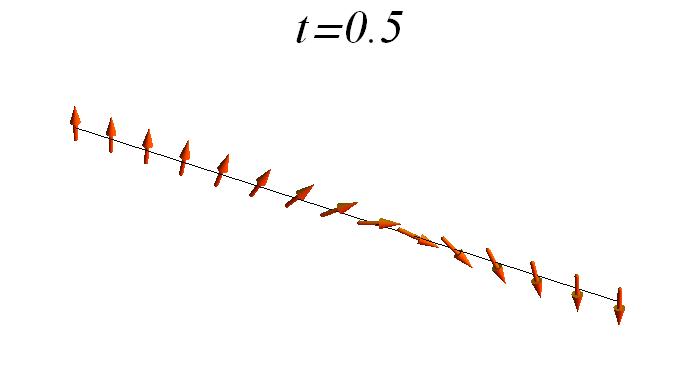}
\includegraphics[width=0.45\textwidth]{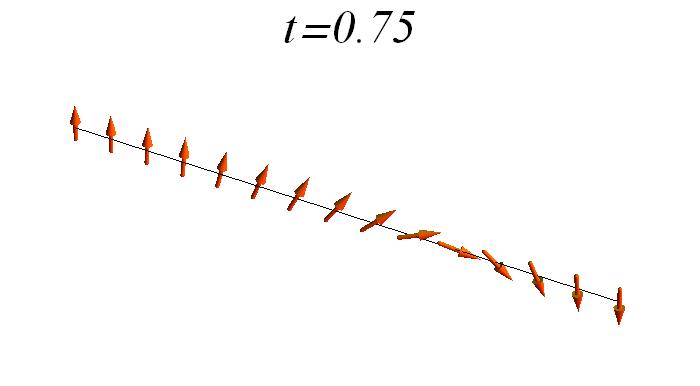}
\includegraphics[width=0.45\textwidth]{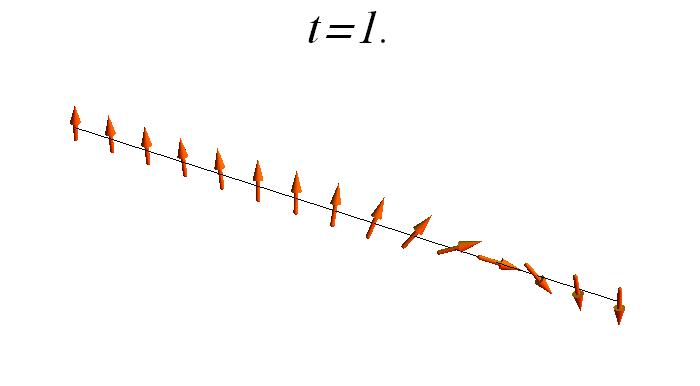}
\caption{\label{fig3} Time evolution of a domain wall of 15 spins with parameters 
$J=1,J_{ex}=0.5,V=1$. 
At the initial time ($t=0$) the domain wall magnetic moments are contained in the
$x-z$ plane. As time evolves the location of the center of the DW shifts to the right.
}
\end{figure*}

\begin{figure*}
\includegraphics[width=0.45\textwidth]{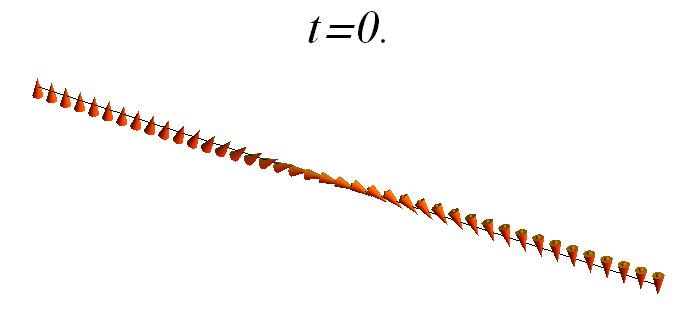}
\includegraphics[width=0.45\textwidth]{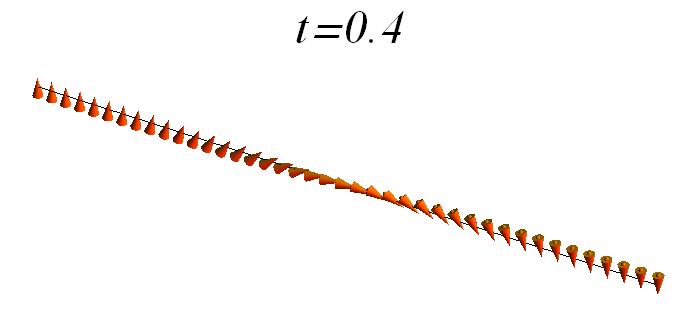}
\includegraphics[width=0.45\textwidth]{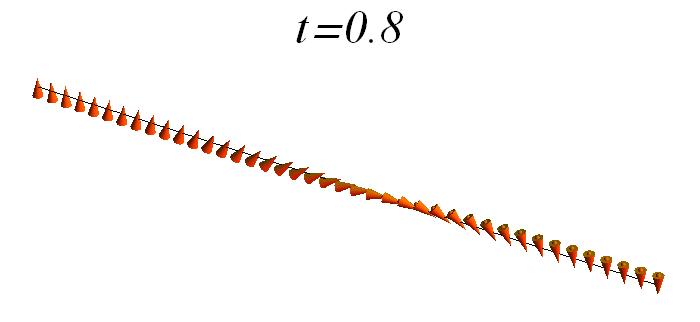}
\includegraphics[width=0.45\textwidth]{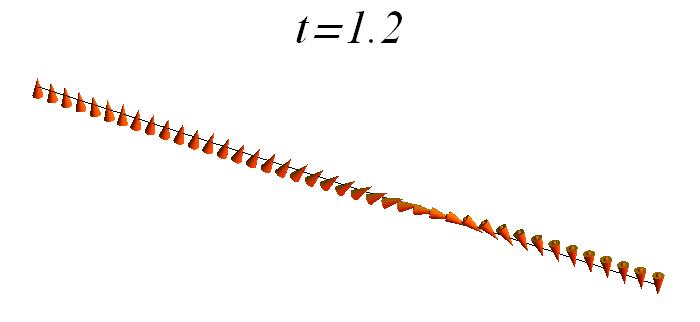}
\includegraphics[width=0.45\textwidth]{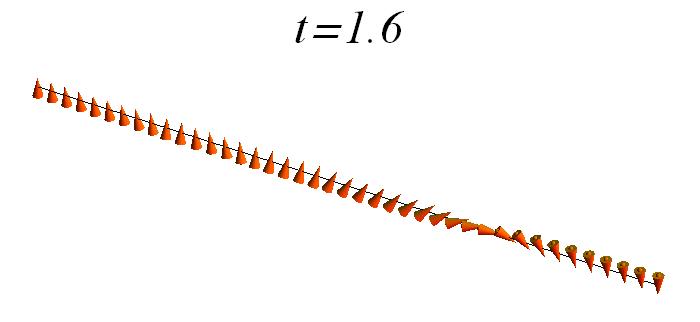}
\includegraphics[width=0.45\textwidth]{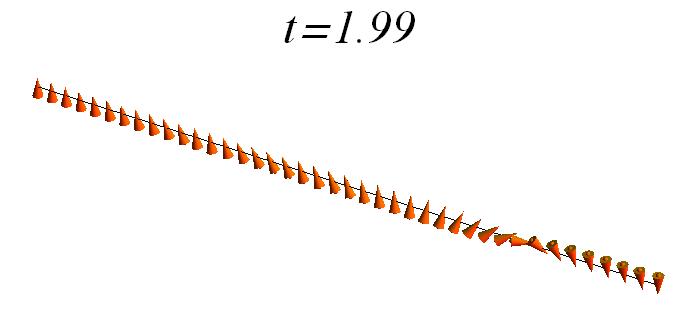}
\caption{\label{fig4} Time evolution of a domain wall of 41 spins with parameters 
$J=1,J_{ex}=0.5,V=1$. 
}
\end{figure*}

Consider now that the spin current is turned on.
The LLG equation for the motion of a magnetic moment $\bm{S}$ in the presence of an
effective magnetic
field $\tilde{\bm{H}}$ and subject to a spin current $\bm{j}^s$ (Slonczewski term \cite{slonczewski}) 
giving origin to a spin torque $\bm{\tau}$ is written as
\be
\frac{d \bm{S}}{dt} = -\gamma \bm{S} \times \tilde{\bm{H}} + \bm{\tau} 
+ \frac{\alpha}{S} \bm{S} \times \frac{d \bm{S}}{dt}
\ee
Here $S$ is the length of the magnetic moment, $\gamma$ is the gyromagnetic ratio 
and $\alpha$ is the damping constant.
The magnetic field $\tilde{\bm{H}}$ is a generic notation for an effective field defined as a
derivative of the energy with respect to the magnetization (spin). 
Due to the spin torque the magnetic
moments in general evolve out of the plane and therefore we assumed the system has an anisotropy term that favors
the plane. Also, the interaction between the local magnetic moments leads to a term 
that contributes to the effective magnetic field. 
The effective magnetic field can be written as
\be
\tilde{\bm{H}} = \bm{H}^y + \bm{H}^{ex}
\ee 
The contributions of these two terms to the effective magnetic field are
\be
\bm{H}^y = -k_y S_y \hat{\bm{y}}
\ee
and
\be
H_{\beta}^{ex}(n)=J_{ex} S_{\beta}(n+1)
\ee
where $\beta=x,y$.

Taking the cross product of the LLG equation with the magnetic moment and using
that its length is fixed ($\bm{S} \cdot d\bm{S}/dt=0$), the LLG can be reformulated as
\bea
\frac{d \bm{S}}{dt} &=& -\gamma_L \bm{S} \times \tilde{\bm{H}} + \frac{\alpha \gamma_L}{S} \bm{S} \times
\left( \bm{S} \times \tilde{\bm{H}} \right) \nonumber \\
&+& \frac{1}{1+\alpha^2} \bm{\tau} + \frac{\alpha}{1+\alpha^2} \frac{1}{S} \bm{S} \times \bm{\tau}
\eea
where
\be
\gamma_L = \frac{\gamma}{1+\alpha^2}
\ee
In terms of the spherical angles $\theta$, $\varphi$ it is written as
\bea
\frac{d \theta}{dt} &=& \gamma_L \left( \tilde{H}_{\varphi} + \alpha \tilde{H}_{\theta} \right) +
\frac{1}{S} \frac{1}{1+\alpha^2} \tau_{\theta} - \frac{1}{S} \frac{\alpha}{1+\alpha^2} \tau_{\varphi}
\nonumber \\
\sin \theta \frac{d \varphi}{dt} &=& -\gamma_L \left( \tilde{H}_{\theta} -\alpha 
\tilde{H}_{\varphi} \right)
+\frac{1}{S} \frac{1}{1+\alpha^2} \tau_{\varphi} + 
\frac{1}{S} \frac{\alpha}{1+\alpha^2} \tau_{\theta}
\nonumber \\
& & 
\eea
where $\tilde{H}_{\theta}$, $\tau_{\theta}$ and $\tilde{H}_{\varphi}$, $\tau_{\varphi}$ 
are the spherical components of the vectors $\tilde{\bm{H}}$ and $\bm{\tau}$, respectively.
The anisotropy and the exchange term imply extra terms in the LLG equations of the form
\bea
H_{\theta}^y &=& -k_y S \sin \theta \cos \theta \sin^2 \varphi \nonumber \\
H_{\varphi}^y &=& -k_y S \sin^2 \theta \sin \varphi \cos \varphi
\eea
and
\bea
H_{\theta}^{ex} &=& J_{ex} \left[ \cos \theta_n \sin \theta_{n+1} \cos \left( \varphi_n-\varphi_{n+1} \right) -
\sin \theta_n \cos \theta_{n+1} \right] \nonumber \\
H_{\varphi}^{ex} &=& -J_{ex} \left[ \sin \theta_{n+1} \sin \left( \varphi_n -\varphi_{n+1} \right) \right]
\eea

These equations are solved iteratively: for a given domain wall configuration, 
at a certain time, we calculate the
electronic properties, such as the spin torque from Eqs. \ref{tdef},\ref{tdef2}. 
This torque is then used to evolve the magnetic moments to the
following time using the LLG equation. The LLG equations change the orientation of the spins. 
This new configuration is then used to calculate the new spin torque and so on. 
To solve the LLG equations we used a
standard second order Runge-Kutta method or the Heun method, yielding similar results.

\section{Evolution of the domain wall}

\begin{figure*}
\includegraphics[width=0.8\textwidth]{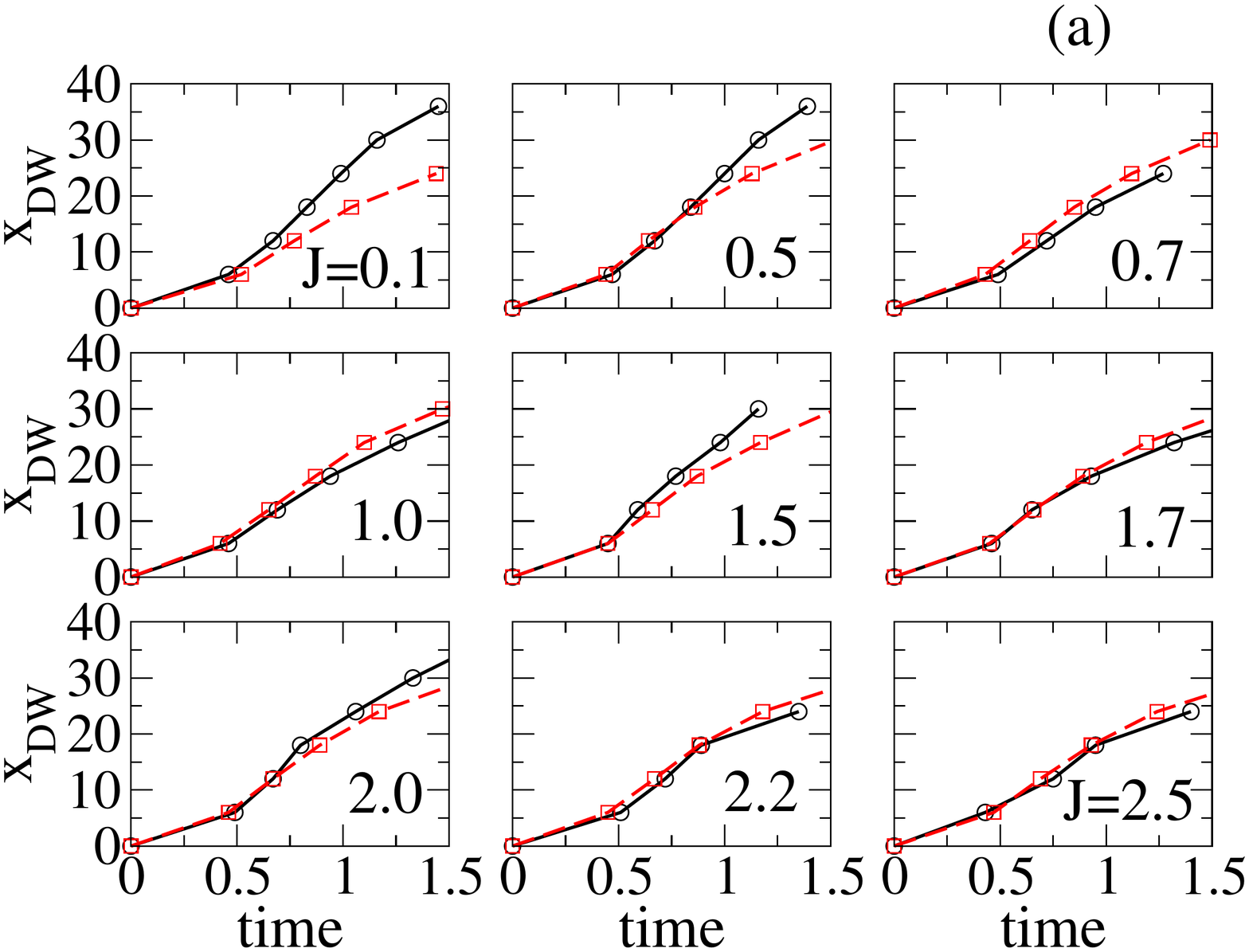}
\includegraphics[width=0.8\textwidth]{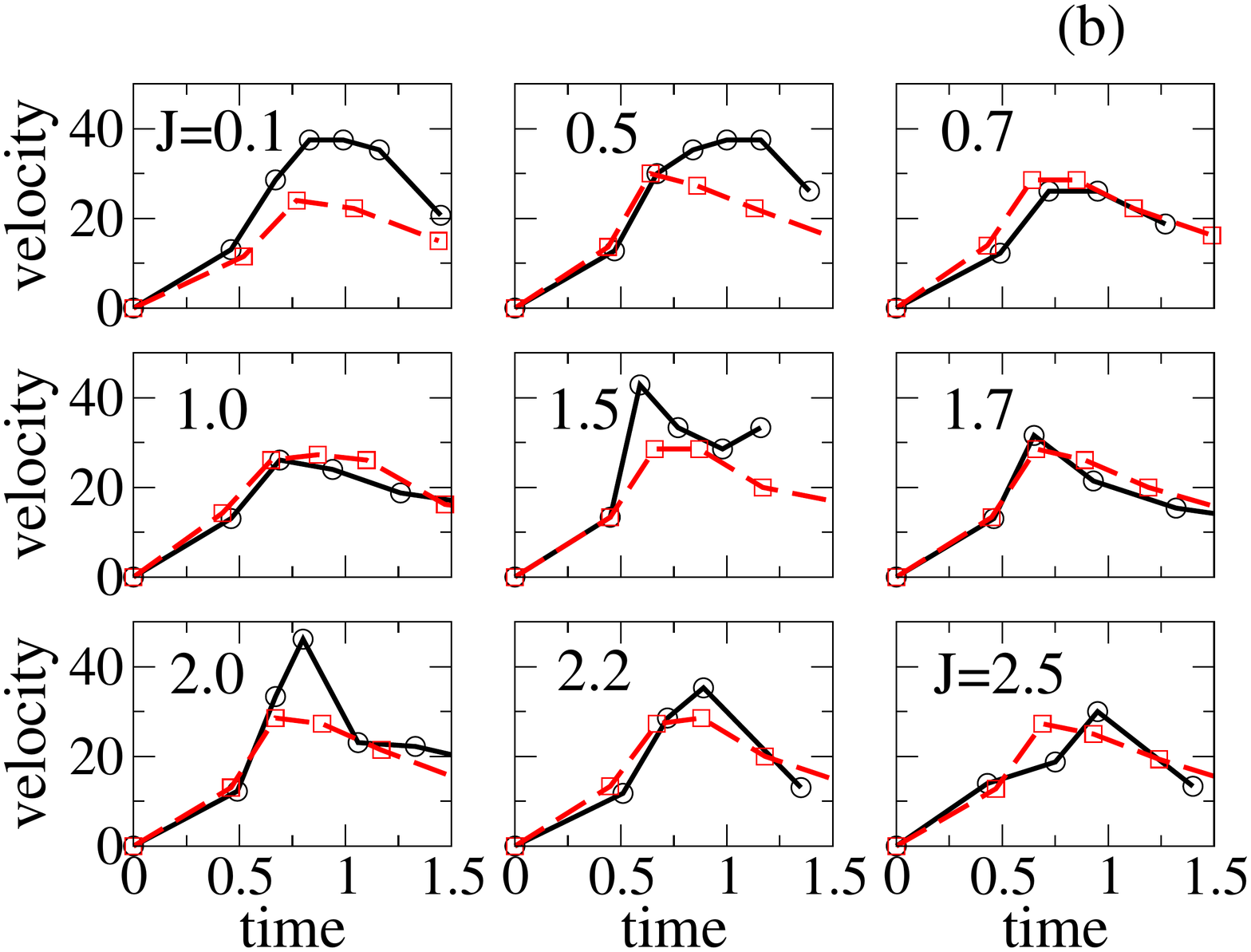}
\caption{\label{fig5} Time dependence of the (a) distance between the center of the DW, $x_{DW}$, and the midpoint of the SC 
and (b) velocity
of the DW for different values of $J$ from $J=0.1, \cdots, 2.5$. 
In black (full lines) are the results
for the superconductor and in red (dashed lines) the results for the normal system.
}
\end{figure*}

\begin{figure*}
\includegraphics[width=0.8\textwidth]{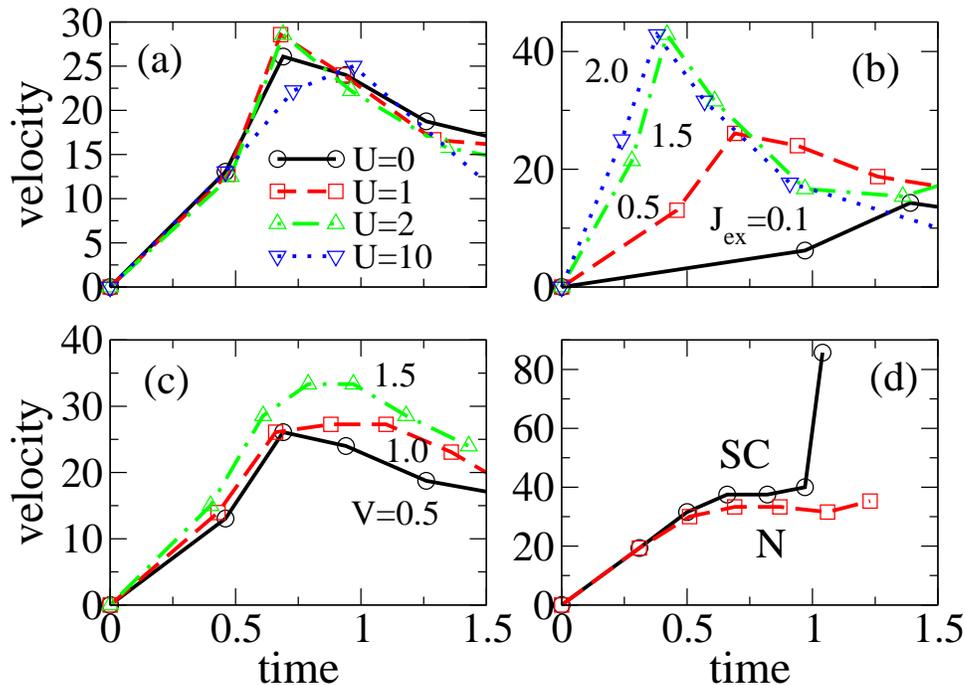}
\caption{\label{fig6} Dependence of the domain wall velocity with the various
parameters. We consider the dependence with (a) the interface disorder, $U$, (b) the exchange
between the spins, $J_{ex}$ and (c) the applied potential, $V$. In panel
(d) and we compare the velocity
for the superconductor and the normal metal for the parameters  
$J=2,J_{ex}=1.5,V=0.5,\alpha=0.5$.
}
\end{figure*}
We start with a domain wall that interpolates between opposite 
exchange fields in the ferromagnetic regions.
Initially the magnetic moments form a N\`eel-type domain wall in the $x-z$ plane. 
As time evolves the domain wall moves to the right due to the spin torque
exerted by the spin polarized current. The spin torque has in-plane and out
of plane components that rotate the magnetic moments. The combined effect
of the spin torque, the exchange coupling between the magnetic moments,
the magnetic anisotropy and the damping term determines the time evolution
of the spins. The magnetic moments tend to rotate to align with the spins
on the left-hand side (leading to the intended displacement of the center
of the DW) but also tend to move out of the $x-z$ plane (this effect is
counteracted by the magnetic anisotropy and by the damping term).

In Figs. \ref{fig3} and \ref{fig4} we show the time
evolution of the domain wall for different parameters.
There are several parameters that need to be specified which implies a very large
parameter space. As a standard case shown in Fig. \ref{fig3} we take $J=1,J_{ex}=0.5,V=1,k_y=4,
\alpha=0.02,U=0,\gamma=2.2,h=0.2$. 
The parameter $h$ is the exchange magnetic field in the ferromagnetic regions. We take $h=0.2$ in the left
F and $h=-0.2$ in the right F.

Here we are interested in studying
the early time evolution, up to the moment the domain wall has come close to the edge of the spin configuration.
In particular, we are interested in comparing the time evolution using a supercurrent with the one obtained
from a standard normal metal. In Fig. \ref{fig5}a we show the displacement of the center
of the DW as a function of time for various parameters.
We define $x_{DW}$ as the distance between the center of the DW and the
midpoint of the SC.

\begin{figure*}
\includegraphics[width=0.45\textwidth]{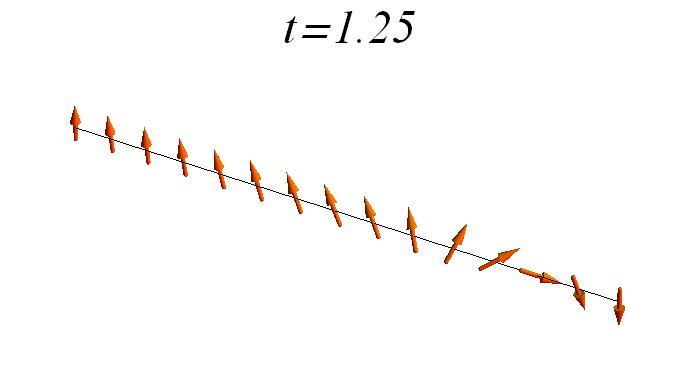}
\includegraphics[width=0.45\textwidth]{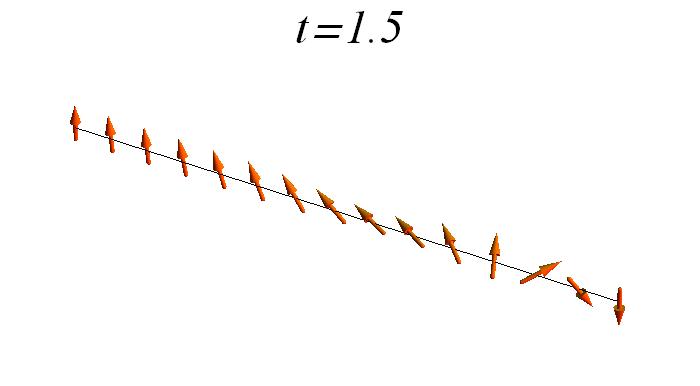}
\includegraphics[width=0.45\textwidth]{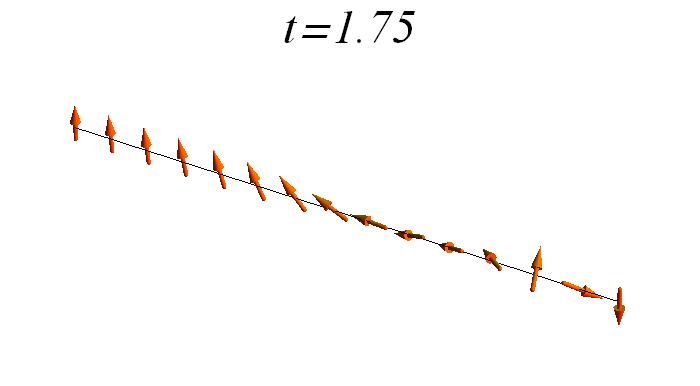}
\includegraphics[width=0.45\textwidth]{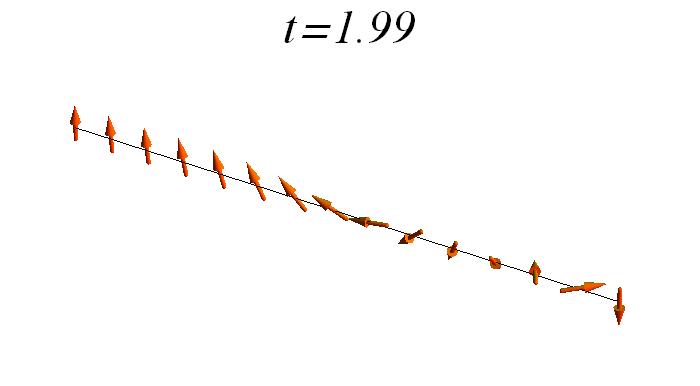}
\caption{\label{fig7} Time evolution of domain wall of 15 spins and 
$J=1,J_{ex}=0.5,V=1$ for later times. 
For the parameters considered here the DW center point reaches the right hand
side border and the remaining non-aligned magnetic moments move out of the $x-z$ plane.
The boundary condition imposed by the right hand side ferromagnet fixes the orientation
of the last magnetic moment ($t=1.25$). At later times $t=1.5,1.75$ the DW starts to
distort and several spins rotate out of the original plane.
}
\end{figure*}

\begin{figure*}
\includegraphics[width=0.45\textwidth]{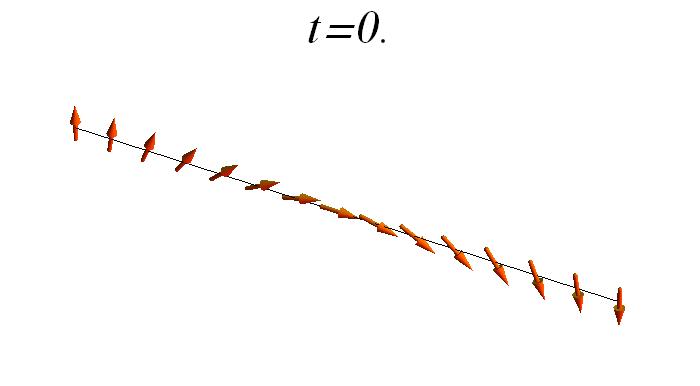}
\includegraphics[width=0.45\textwidth]{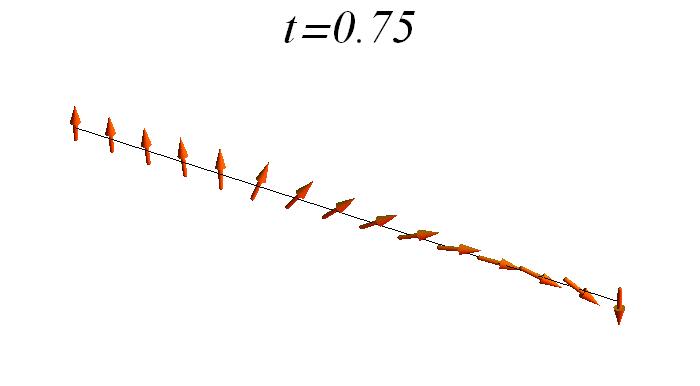}
\includegraphics[width=0.45\textwidth]{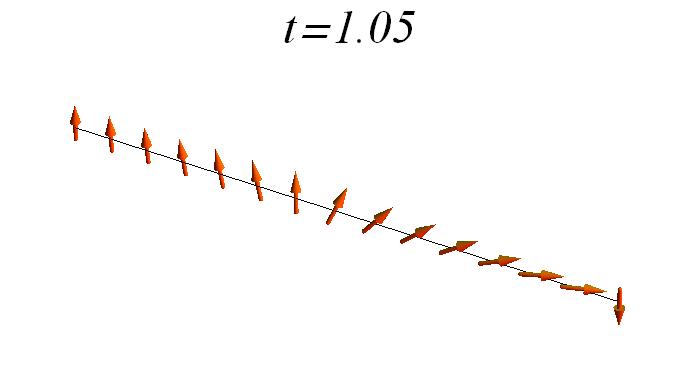}
\includegraphics[width=0.45\textwidth]{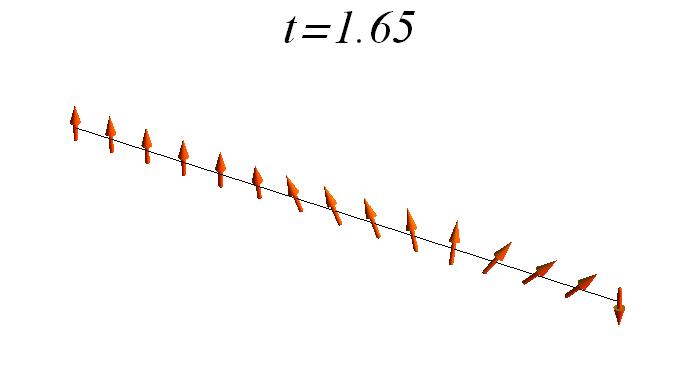}
\caption{\label{fig8} Time evolution of domain wall of 15 spins and 
$J=2,J_{ex}=1.5,V=0.5,\alpha=0.5$. Increasing the couplings between
the magnetic moments and their coupling to the electron spin density,
and the Gilbert damping, the domain wall distortion is smaller,
even though some spins start to move out of the initial plane when
the center of the DW approaches the right hand boundary ($t=1.65$).
}
\end{figure*}

To better quantify the motion of the domain wall we calculate the velocity shown
in Fig. \ref{fig5}b. 
In the initial configuration the spins on the left hand side have an orientation
such that $\theta<\pi/2$ while on the right hand side $\theta>\pi/2$. As the domain wall
moves to the right, the orientation of spins on the right hand side will successively cross
$\pi/2$.
We define the center of the DW to be the spin with $\theta=\pi/2$. The velocity of the DW
is $(d x_{DW}/dt)$.. 
This is shown in Fig. \ref{fig5}b for different values of $J$, where we 
compare the velocities as a function
of time for the superconductor and a normal system. This last case is 
simply obtained turning off the superconducting pairing. 

Typically, at early times, the velocity is small. It increases over time and then
starts to saturate. 
This is related to either basically all spins have already flipped or the DW starts to get
distorted and the last spins do not flip at all. 
As discussed in Ref. \cite{dugaev2} the spin torque has in general a component
that leads to the motion of the magnetic moments out of the original DW plane.
This term implies an acceleration of the motion of the domain wall that leads
to an increased velocity. The velocity then saturates both due to the presence
of friction originated by the damping term, and by shear finite size effects
due to the approach to the boundary of the system. 
As a function of the coupling between the magnetic moments and the electron spin density, $J$,
we see various regimes. At small $J$ the DW moves faster for the superconductor, 
between $J=0.7-1.2$ the normal
metal is more efficient, and then it changes from one to the other. 
It may seem therefore that the velocity may be larger if the magnetization is smaller. 
However, there are regimes
for larger $J$ for which the velocity is larger in the superconducting phase, 
such as $J=1.5,2$. Also, one would
expect that the velocity should be larger if the spin torque is larger. 
The values of the velocities are similar for all cases
but for small $J$ and for the case of $J=1.5$ the velocity in the superconductor is particularly larger with
respect to the normal metal. For this particular value of the coupling and for this
system size the velocity does
not seem to saturate before the domain wall gets distorted.

To establish in greater detail the influence of the various parameters we consider a few
cases. Due to the large parameter space we fix the various parameters at the standard case
and change one of the parameters to see its influence. In Fig. \ref{fig6} we change the
interface disorder $U$, the coupling between the magnetic moments, $J_{ex}$, and the external
potential, $V$. As one expects, the velocity is typically larger for a clean interface, or at least
for small interface disorder. The velocity increases as the exchange $J_{ex}$ increases and as
the potential, $V$, increases. Small interface disorder is important to have larger currents,
and larger potential has the same effect. Larger currents lead to larger spin torques.
The effect of the exchange interaction is also clear because it tends to collectively 
move the spins.

From experimental results one knows that the domain walls either stop, distort or
change into vortex configurations \cite{nat1,nat2}, and the process has to be restarted. 
This is not related however to the finite size effects of the wire. In any case we were
mostly interested in describing the early times displacement of the DW.

Since the system sizes considered are finite, the motion of
the domain wall is conditioned by the end of the system. 
Typically the domain wall moves from left to right
and, when the spins close to the edge move appreciably, there is an important distortion 
of the domain wall since the magnetic moments rotate out of the original plane. 
This occurs near the right edge of the system but often it propagates to the left. In some
cases it destroys the domain wall on the right hand side and the spins on the left hand side may
start to precess and move out of the plane. Clearly, these situations are not what is intended since
we want to move the center of the domain wall but maintain the spins aligned in the original
plane (defined as $x-z$). This is illustrated in Fig. \ref{fig7}. 
Beyond the point for which the center of the DW approaches the right boundary
the spins on the right hand side start to disorder 
(in the sense they loose the alignment with the neighboring spins) and the spins on
the left hand side start to move out of the plane. 

This problem may be avoided (for not very long times) by changing the parameters. We
present in Fig. \ref{fig8}
the time evolution of the same domain wall with 15 spins (recall that the system size is 100 sites and therefore
15 spins is rather dilute) where we have changed some of the parameters to a set with $J=2,J_{ex}=1.5,V=0.5,\alpha=0.5$.
The motion of the domain wall is now much better controlled, at least up to a time of the order of $t=2$.
Clearly, when all the spins have been flipped one should turn the current off. We have found that for these
parameters it is important to increase the coupling between the magnetic moments and the spins of the
conduction electrons, to increase the coupling between the magnetic moments and 
to increase the damping term,
so that the motion of the magnetic moments does not lead to a large distortion
of the DW. 
As shown in Fig. \ref{fig6}d in this case the SC velocity is considerably larger
than the velocity in the normal phase.

Another way to make the evolution of the domain wall well controlled is to increase the density
of the magnetic moments as shown in Fig. \ref{fig4}. 
The evolution is slower but well controlled
up to $t=2$, for which the domain wall has not yet reached the border. 

\section{Conclusions}

Following previous works where we considered the possibility of the effects of magnetic
impurities, immersed or in the vicinity of a conventional superconductor, paying attention
to both the effects of the impurities on the superconductor and of the superconductor
on the magnetic impurities, we have solved the LLG equations for the motion of a domain wall
due to the passage of a spin polarized current through a heterostructure of the type FSF.
The new aspect is that the magnetic moments are distributed in the S region.
Consistently with previous results, where the spin torque exerted on the magnetic
moments may be larger in the superconductor as compared to a normal metal, we have found
that the velocity of the motion of the domain wall in the superconducting phase may be enhanced
with respect to the normal phase.

\section{Acknowledgments}

We acknowledge discussions with V. Dugaev, P. Horley, J. Barnas and thank P. Ribeiro
for help with graphical aspects. 
This research was partially supported 
by FCT-Portugal through grant PTDC/FIS/70843/2006.



\begin{thebibliography}{}

\bibitem{rmp1} I. Zutic, J. Fabian and S. Das Sarma, Rev. Mod. Phys. {\bf 76}, 323 (2004).

\bibitem{fert} A. Fert, Thin Solid Films {\bf 517}, 2 (2008).

\bibitem{zabel} H. Zabel, Superlattices and Microstructures, {\bf 46}, 541 (2009).

\bibitem{nam} C. Nam, Y.M. Jang, K.S. Lee, S. K. Lee, T.W. Kim, B.K. Cho,  J. Magn.
Magn. Mater.  {\bf 310}, 2023 (2007).

\bibitem{ralph} D.C. Ralph and M.D. Stiles, J. Magn. Magn. Mater. {\bf 320}, 1190 (2008).

\bibitem{slonczewski} J.C. Slonczewski, J. Magn. Magn. Mater. {\bf 159}, L1 (1996).

\bibitem{berger} L. Berger, J. Appl. Phys. {\bf 71}, 2721 (1992); Phys. Rev. B {\bf 54}, 9353 (1996).

\bibitem{wang} K.Y. Wang, A.C. Irvine, R.P. Campion, C.T. Foxon, J. Wunderlich, D. A.
Williams, B.L. Gallagher, J. Magn. Magn. Mater.  {\bf 321}, 971 (2009).


\bibitem{tatara} S. Takagi and G. Tatara, Phys. Rev. B {\bf 54}, 9920 (1996); G. Tatara and H. Kohno,
Phys. Rev. Lett. {\bf 92}, 086601 (2004); G. Tatara, N. Vernier and J. Ferre, Appl. Phys. Lett. {\bf 86},
252509 (2005).

\bibitem{dugaev1} V.K. Dugaev, J. Berakdar and J. Barnas, Phys. Rev. B {\bf 68}, 104434
(2003).

\bibitem{dugaev2} V.K. Dugaev, V.R. Vieira, P.D. Sacramento, J. Barnas, 
M.A.N. Ara\'ujo and J. Berakdar, 
Phys. Rev. B {\bf 74}, 054403 (2006).

\bibitem{berger1} L. Berger, J. Appl. Phys. {\bf 3}, 2156 (1978); 2137 (1979); P.P. Freitas and
L. Berger, J. Appl. Phys. {\bf 57}, 1266 (1985); C.-Y. Hung, L. Berger, J. Appl. Phys. {\bf 63},
4276 (1988).

\bibitem{dws} J. Grollier et al., Appl. Phys. Lett. {\bf 83}, 509 (2003);
N. Vernier, D. A. Allwood, D. Atkinson, M. D. Cooke and R. P. Cowburn, Europhys. Lett.
{\bf 65}, 526 (2004);
A. Yamaguchi et al., Phys. Rev. Lett. {\bf 92}, 077205 (2004);
M. Klaui et al., Phys. Rev. Lett. {\bf 94}, 106601 (2005);
M. Klaui et al., Phys. Rev. Lett. {\bf 95}, 026601 (2005);
M. Yamanouchi, D. Chiba, F. Matsukura and H. Ohno, Nature {\bf 428}, 539 (2004);
D. Ravelosona, D. Lacour, J. A. Katine, B. D. Terris and C. Chappert, Phys. Rev.
Lett. {\bf 95}, 117203 (2005).


\bibitem{linders} J. Linder, T. Yokoyama, A. Sudbo, Phys. Rev. B {\bf 79}, 054523 (2009);
J. Linder, T. Yojoyama and A. Sudbo, Phys. Rev. B {\bf 79}, 224504 (2009);
J. Linder, A. Sudbo, arXiv:1004.5124 (2010).

\bibitem{satori} K. Satori, H. Shiba, O. Sakai and Y. Shimizu, J. Phys. Soc. Jpn. 
{\bf 61}, 3239 (1992); P. Schlottmann, Sol. Stat. Comm. {\bf 16}, 1297 (1975).

\bibitem{sakurai} A. Sakurai, Prog. Theor. Phys. {\bf 44}, 1472 (1970); A.V. Balatsky, I. Vekhter, J.-X. Zhu,
Rev. Mod. Phys. {\bf 78}, 373 (2006).

\bibitem{us} P.D. Sacramento, V.K. Dugaev and V.R. Vieira, Phys. Rev. B {\bf 76}, 014512 (2007).

\bibitem{abrikosov} A.A. Abrikosov and L.P. Gorkov, Zh. Eksp. Teor. Fiz. {\bf 39}, 178 (1960)
[Sov. Phys. JETP {\bf 12}, 1243 (1961)].

\bibitem{rapid} P. D. Sacramento, V. K. Dugaev and V. R. Vieira, Phys. Rev. B {\bf 76}, 020510(R) (2007).

\bibitem{jpcm2010} P.D. Sacramento, V.K. Dugaev, V.R. Vieira and M.A.N. Ara\'ujo,
J. Phys. Cond. Matt. {\bf 22}, 025701 (2010). 

\bibitem{torquesc} P. D. Sacramento and M. A. N. Ara\'ujo, Europ. Phys. J. B {\bf 76},
251 (2010).

\bibitem{beach} G.S.D. Beach, M. Tsoi and J.L. Erskine, J. Mag. Mag. Mat. {\bf 320}, 1272 (2008).

\bibitem{yaroslav} Y. Tserkovnyak, A. Brataas and G.E.W. Bauer, J. Mag. Mag. Mat. {\bf 320}, 1282 (2008).

\bibitem{ting} J.-X. Zhu and C.S. Ting, Phys. Rev. B {\bf 61}, 1456 (2000).

\bibitem{degennes} P. G. de Gennes, {\it Superconductivity of metals and alloy}, (Reading,
MA: Addison-Wesley Publishing Company, Inc.) 1989.

\bibitem{blonder} G.E. Blonder, M. Tinkham and T.M. Klapwijk, Phys. Rev. B {\bf 25}, 4515 (1982).

\bibitem{maekawa} T. Yamashita, H. Imamura, S. Takahashi and S. Maekawa, Phys. Rev. B {\bf 67},
094515 (2003).

\bibitem{lambert} C.J. Lambert, J. Phys. Cond. Matt. {\bf 3}, 6579 (1991).

\bibitem{LLG} T. L. Gilbert, IEEE Trans. Mag. 40, 3443 (2004);
L. D. Landau and E. M. Lifshitz, Phys. Z. Sowietunion 8, 153 (1935);
J. Xiao, A. Zangwill and M. D. Stiles, Phys. Rev. B  72, 014446 (2005);
P. P. Horley, V. R. Vieira, P. M. Gorley, V. K. Dugaev, J. Berakdar and
J. Barnas, Phys. Rev. B 78, 054417 (2008);
P. P. Horley, V. R. Vieira, P. M. Gorley, V. K. Dugaev,
J. Barnas, J. Magn. Magn. Mater. 322, 1434-1437 (2010).

\bibitem{entanglement} P. D. Sacramento, P. Nogueira, V. R. Vieira and V. K. Dugaev, Phys. Rev. B {\bf 76}, 184517 (2007).

\bibitem{fidelity} N. Paunkovic, P. D. Sacramento, P. Nogueira, V. R. Vieira and V. K. Dugaev, Phys. Rev. A
{\bf 77}, 052302 (2008).

\bibitem{loff} P. Fulde and R. A. Ferrell, Phys. Rev. {\bf 135}, A550 (1964); A. I. Larkin and
Y. N. Ovchinnikov, Zh. Eksp. Teor. Fiz. {\bf 47}, 1136 (1964) [Sov. Phys. JETP 20, 762
(1965)].

\bibitem{nat1} M. Hayashi, L. Thomas, C. Rettner, R. Moriya, S. S. P.
Parkin, Nature {\bf 443}, 197 (2006)

\bibitem{nat2} L. Thomas, M. Hayashi, X. Jiang, R. Moriya, C. Rettner,
S. S. P. Parkin, Nature Physics {\bf 3}, 21 (2007).

\end{thebibliography}
\end{document}